\documentclass[10pt,twocolumn,final]{article}
\pdfoutput=1

\usepackage[top=2cm, left=1.5cm, bottom=2cm, right=1.3cm]{geometry}

\usepackage{amsmath}
\usepackage{amssymb}
\usepackage{latexsym}

\usepackage{graphicx}
\usepackage{balance}

\usepackage[square,numbers,sort&compress]{natbib}

\usepackage{fancyhdr}
\makeatother
\usepackage{url}
\usepackage{xcolor}
\usepackage{hyperref}
\usepackage{pdfpages}

\usepackage{amssymb}
\usepackage{amsmath}
\usepackage{bm}

\newcommand{\spipe}{{\ } | {\ }}
\newcommand{\sssize}[1]{\text{\sffamily{\tiny{#1}}}}

\newcommand{\se}[1]{\textit{SE(#1)}}
\newcommand{\so}[1]{\textit{SO(#1)}}

\begin{document}

\twocolumn[
  \begin{@twocolumnfalse}

\title{
Orientation-Disentangled Unsupervised Representation Learning\newline
for Computational Pathology
}
\date{}

\vspace*{-50pt}
\begin{minipage}{\textwidth}
\centering
\author{
Maxime W. Lafarge,
Josien P.W. Pluim,
Mitko Veta}

\end{minipage}

\maketitle

\vspace*{-20pt}\hspace*{20px}
\begin{minipage}{0.9\textwidth}
\begin{center}
\footnotesize
\textit{Department of Biomedical Engineering, Eindhoven University of Technology, Eindhoven, The Netherlands} 
\end{center}
\end{minipage}

\vspace*{40pt}
\begin{abstract}
Unsupervised learning enables modeling complex images without the need for annotations.
The representation learned by such models can facilitate any subsequent analysis of large image datasets.

However, some generative factors that cause irrelevant variations in images can potentially get entangled in such a learned representation causing the risk of {negatively affecting} any subsequent use.
The orientation of imaged objects, for instance, is often arbitrary/irrelevant, thus it can be desired to learn a representation in which the orientation information is disentangled from all other factors.

Here, we propose to extend the Variational Auto-Encoder framework by leveraging the group structure of rotation-equivariant convolutional networks to learn orientation-wise disentangled generative factors of histopathology images.
This way, we enforce a novel partitioning of the latent space, such that oriented and isotropic components get separated.

We evaluated this structured representation on a dataset that consists of tissue regions for which nuclear pleomorphism and mitotic activity was assessed by expert pathologists.
We show that the trained models efficiently disentangle the inherent orientation information of single-cell images.
In comparison to classical approaches, the resulting aggregated representation of sub-populations of cells produces higher performances in subsequent tasks.
\end{abstract}
\vspace*{50pt}

  \end{@twocolumnfalse}
]

\section{Introduction}
\label{introduction}

Dimensionality reduction is an efficient strategy to facilitate the analysis of large image datasets by representing individual images by a small set of informative variables, {which} can be used in place of the original images.
Unsupervised learning methods can be used to obtain such an informative latent representation from a given dataset without the need for expert annotations.
For this purpose, popular unsupervised learning frameworks such as the Variational Auto-Encoder \citep{kingma2013vae} or flow-based approaches \citep{rezende2015variational} can be used to model a joint distribution between an image dataset and a set of latent generative factors.
As these frameworks provide a posterior distribution over a space of latent variables, they enable the estimation of the latent factors of {new previously unseen images}, that can then be used for any subsequent task.

However, irrelevant factors that affect the appearance of images but are independent of the factors of interest can get \textit{entangled} in the learned representation \citep{louizos2016fair, higgins2017beta, higgins2018definition}.
These irrelevant factors can be treated as \textit{nuisance variables} that affect the learned representation in an unpredictable way \citep{belthangady2019applications}, consequently perturbing any \textit{downstream analysis}\footnote{We refer to \textit{downstream task/analysis} to express any task/analysis performed on an image dataset for which a learned representation is used in place of original images.} performed on a distribution of generative factors.
Therefore, there is a need for \textit{disentangling} such nuisance variables from the informative generative factors of interest.

In computational pathology, these nuisance variables are known to affect the generalization power of machine learning models.
They affect the appearance of the images across slides, scanners and hospitals and can be {associated with} the inevitable variations in tissue slide preparation and scanner-dependent digitization protocols.

In a supervised {learning} context, strategies were developed to filter-out such irrelevant factors from the learned representation; popular methods applied in computational pathology include: staining normalization \citep{ciompi2017sn}, staining/style transfer \citep{bentaieb2017adversarial,gadermayr2018way,debel2019stain}, data augmentation \citep{tellez2018heAugmentation}, domain-adversarial {training} \citep{lafarge2017domain,lafarge2019domain} and rotation-equivariant modeling \citep{bekkers2018roto,lafarge2020roto}.

In this paper, we focus on a specific generative factor that can be considered as a nuisance variable {in some specific tasks}: the orientation of individual image patches.
In digital pathology, the orientation of localized image patches in a dataset of WSIs is arbitrary in the sense that tissue structures are likely to be observed in any orientation, as opposed to natural images or organ-level medical images for which the orientation of the imaged objects is typically not uniformly distributed.

\textbf{We {propose an unsupervised learning framework to model} a partitioned latent space of generative factors {in which} {specific} independent latent variables either code for oriented or non-oriented (isotropic) morphological components of {histopathology} images.}

\paragraph{\textbf{Motivation}}
We identified several points that motivate the development of methods to handle nuisance {variables} for computational pathology in an unsupervised {learning} context:
\begin{itemize}
	\setlength\itemsep{0pt}
	\item[-] Using an informative representation in place of large and complex images can reduce the computational cost and facilitate training of subsequent task-specific models.
In particular, such a representation can be used to directly process Whole Slide Images (WSIs) via patch-based compression \citep{tellez2019neural} or to represent bags that consist of a high number of image patches in two-stage multiple-instance-learning frameworks \citep{ilse2020MILforHIA}.
By removing irrelevant factors from the representation, such existing frameworks can be further improved.

	\item[-] A representation learned without supervision can better conserve the extent of the morphological information of tissue images making it suitable for a wide range of potential downstream tasks.
	This is opposed to using the representation of a supervised model that potentially discards information that is irrelevant for the task for which it was trained, but that might be relevant for other downstream tasks.

	\item[-] Latent variable models equipped with a generative component enable visual inspection of the individual learned factors.
This can support the interpretation of a model, as a tool to gain insights into the morphological factors that are predictive for a given downstream task.
\end{itemize}

\begin{figure}[ht!]
\begin{center}
\includegraphics[width=\columnwidth, trim=15pt 380pt 310pt 5pt, clip]{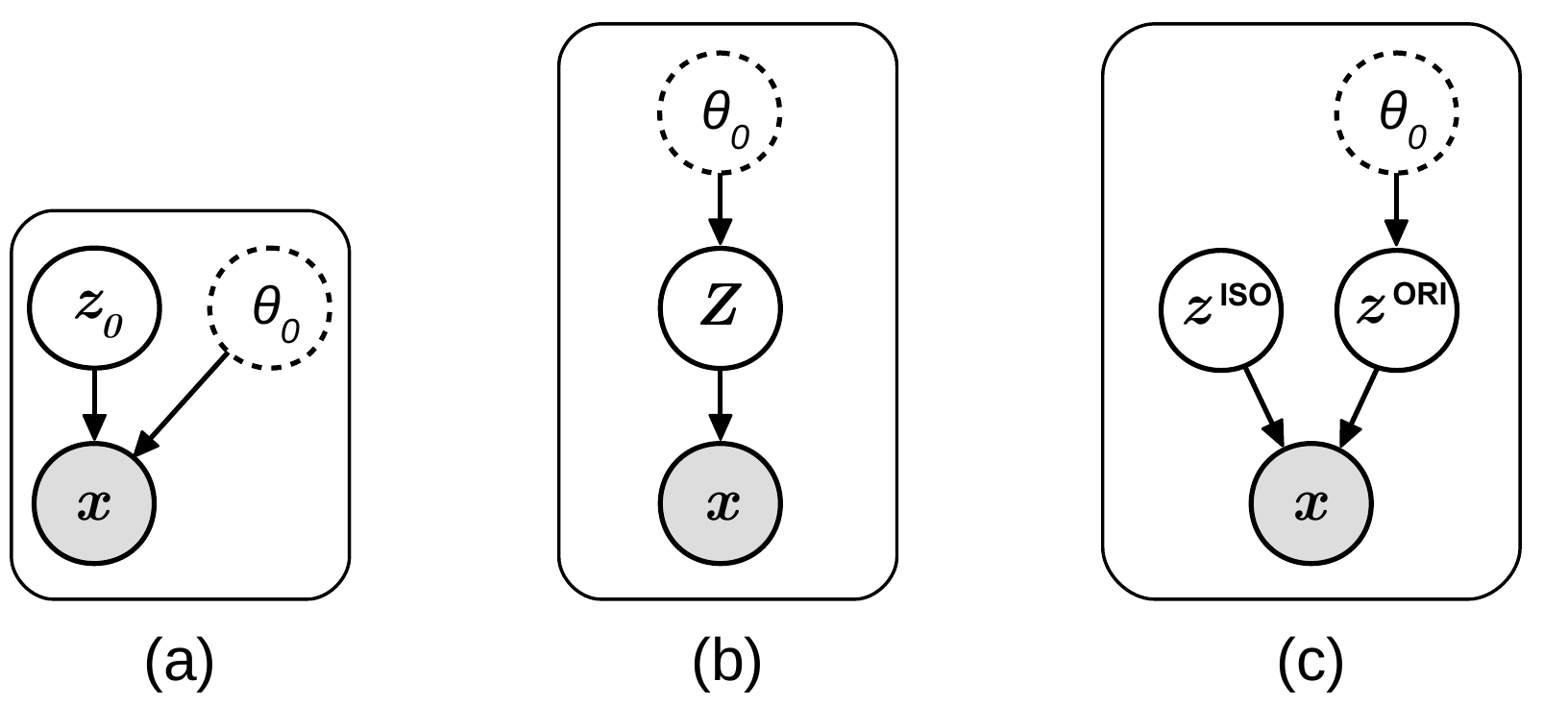}
\caption{\footnotesize
{
Bayesian networks of three latent variable models in which a hidden nuisance variable $\theta_{0}$ is involved.
(a) Classical generative model; $\bm{z}_{0}$ and $\theta_{0}$ are independent sources of the observed images $\bm{x}$.
(b) Chain-structured model; the images $\bm{x}$ are generated from intermediate latent variables $\bm{Z}$ that depend on $\theta_{0}$.
(c) Proposed disentangled model; the images $\bm{x}$ are generated by two independent variables: $\bm{z}^{\sssize{ISO}}$ that is independent of $\theta_{0}$ and $\bm{z}^{\sssize{ORI}}$ that encodes $\theta_{0}$.
}
}
\label{fig:graphicalModels}
\end{center}
\end{figure}

\begin{figure*}[ht!]
\begin{center}
\includegraphics[width=0.8\textwidth, trim=5pt 160pt 220pt 5pt, clip]{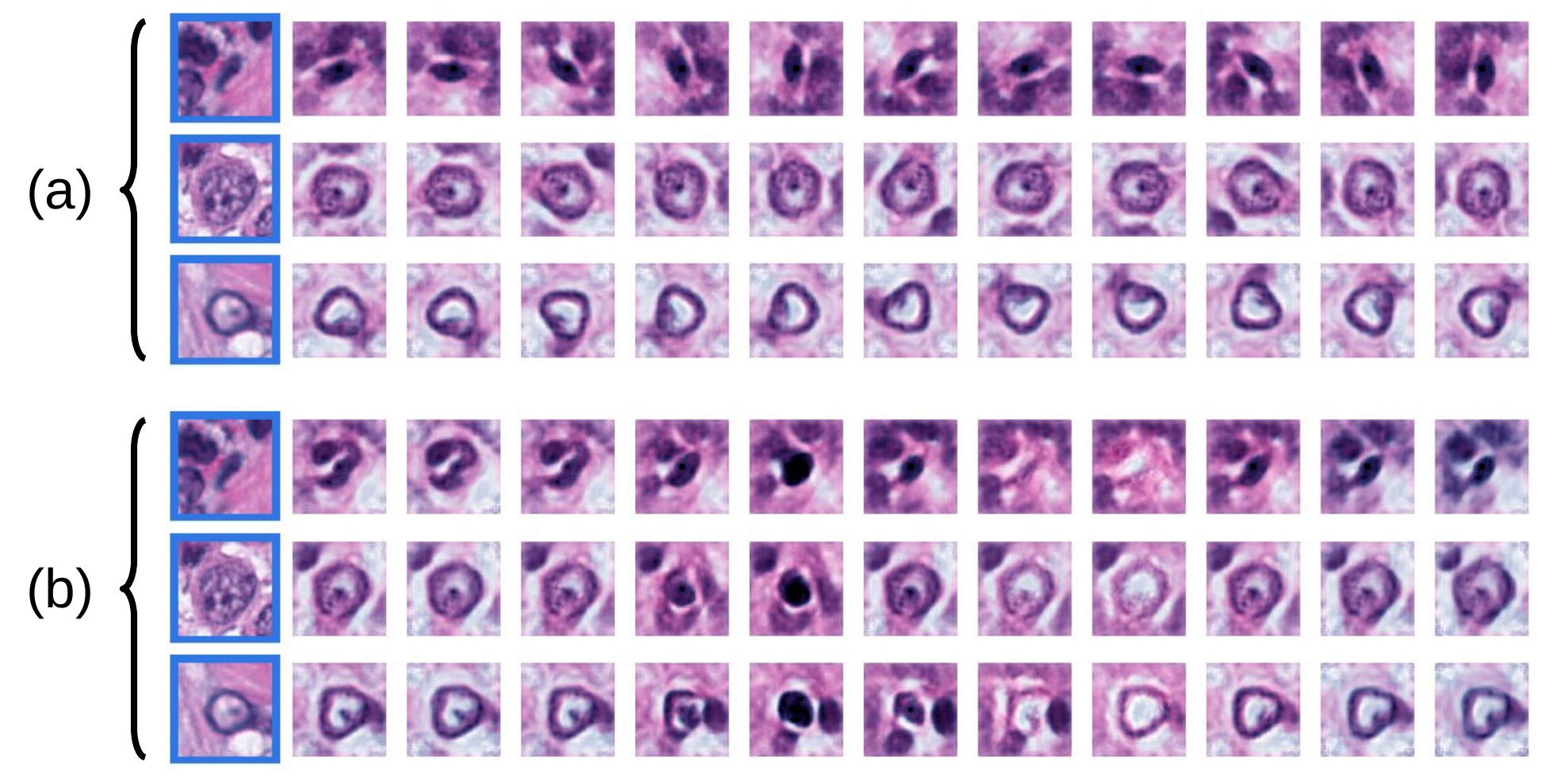}
\caption{\footnotesize{
Generated images using the proposed latent variable model in which images are represented by a set of two types of variables: real-values $\bm{z}^{\sssize{ISO}}$ that code for isotropic components and angle variables $\bm{z}^{\sssize{ORI}}$ that code for oriented components.
{Left-most} images are original and were used to estimate initial values of $\bm{z}^{\sssize{ISO}}$ and $\bm{z}^{\sssize{ORI}}$.
In (a), $\bm{z}^{\sssize{ISO}}$ is kept fixed and the values of $\bm{z}^{\sssize{ORI}}$ are sequentially incremented by a fixed angle measurement (cycle-shifted), causing a spatial rotation of the generated images.
In (b), $\bm{z}^{\sssize{ORI}}$ is kept fixed while the values of $\bm{z}^{\sssize{ISO}}$ are sequentially varied causing \textit{isotropic} morphological changes in the generated images.
}} 
\label{fig:generativeProcess}
\end{center}
\end{figure*}

\paragraph{\textbf{The proposed method}}
To enable such a partitioning of the latent space, we leveraged the structure of \se{2}-group convolutional networks \citep{bekkers2018roto, lafarge2020roto} to build the components of a new VAE that produces {a pair} of rotation-equivariant and rotation-invariant embeddings of image patches.
This means that instead of representing an image by a vector of scalar latent variables ($\bm{z}_{0}$ in Figure.~\ref{fig:graphicalModels}~(a)), the proposed framework learns {in parallel} a vector of real-valued isotropic variables ($\bm{z}^{\sssize{ISO}}$) and a vector of angular orientation variables ($\bm{z}^{\sssize{ORI}}$, see Figure.~\ref{fig:graphicalModels}~(c)) that enable the disentanglement of the orientation information in images.

In Figure.~\ref{fig:generativeProcess}, we illustrated the orientation disentanglement property obtained with the proposed framework, in which the effect of varying each type of generative factor can be observed in the generated examples.
{The resulting structure of the learned representation is also illustrated in Figure.~\ref{fig:generativeInterpolation}, in which we show that the translation {between existing datapoints} in the latent space along the oriented or isotropic dimenisons causes distinctive generative effects.

The independence between the two types of variables is guaranteed by the structure of the \se{2}-CNN-based auto-encoder.
In the spirit of the original VAE framework, we propose an extension of the {objective function} so as to encourage the mutual independence between all the introduced latent variables.

\begin{figure*}[ht!]
\begin{center}
\includegraphics[width=0.8\textwidth, trim=5pt 150pt 200pt 5pt, clip]{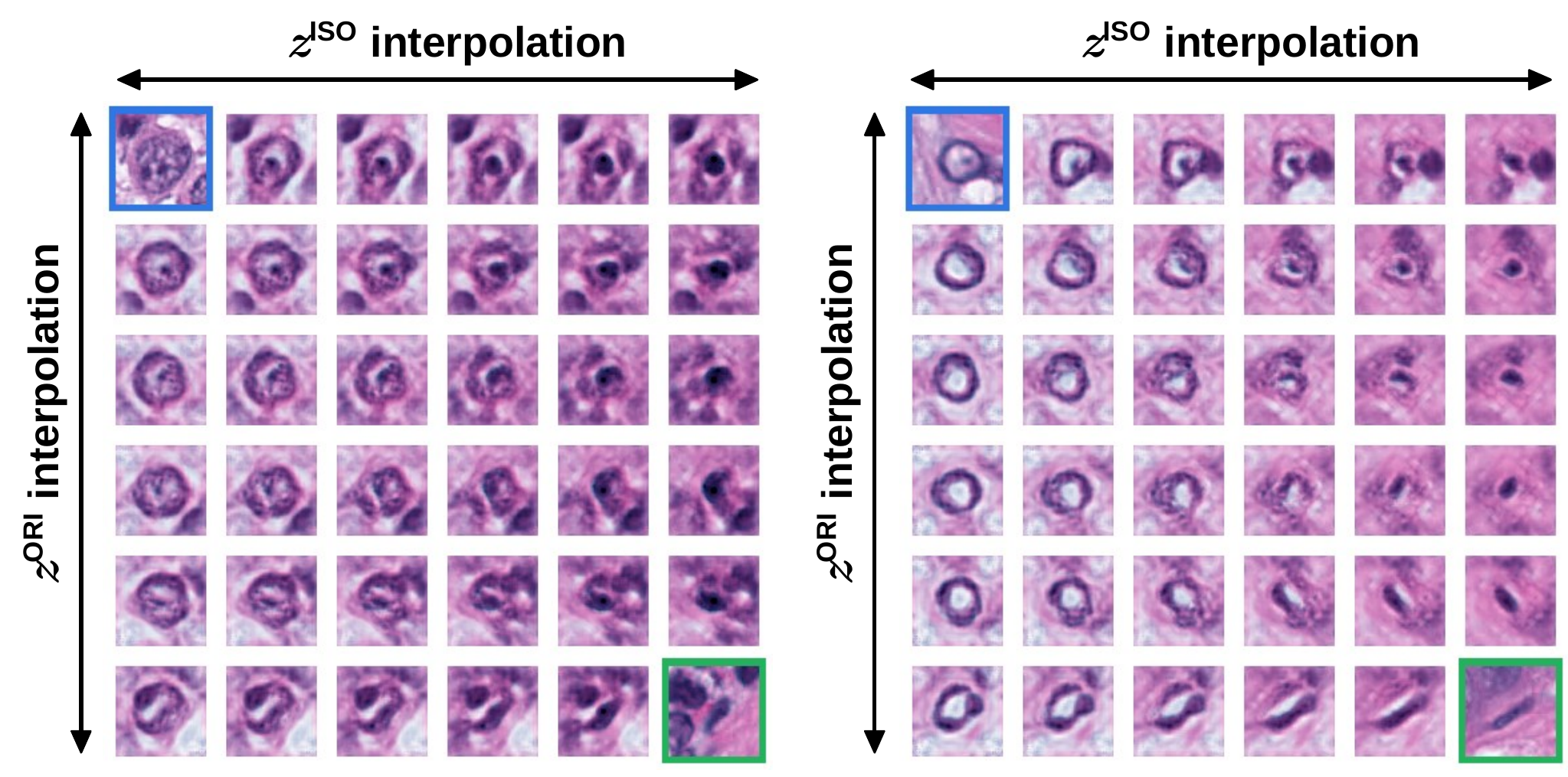}
\caption{\footnotesize{
Generated image from linearly interpolated latent variables: along the horizontal axis the isotropic variables $\bm{z}^{\sssize{ISO}}$ are interpolated, along the vertical axis the angular components $\bm{z}^{\sssize{ORI}}$ are interpolated.
Original images used to estimated the original values of the latent variables are framed in blue (top-left images), and green (bottom-right images).
}} 
\label{fig:generativeInterpolation}
\end{center}
\end{figure*}

We made a comparative analysis of the proposed framework with a baseline VAE and well-established hand-crafted measurements in the context of histopathology image analysis.
We trained and evaluated the models using a dataset of nucleus-centered images extracted from histological cases of $148$ breast cancer patients, exposing a large variability of nuclear morphology to be learned. 
We evaluated the quality of the unsupervised learned representation by training simple logistic regression models on downstream classification tasks. We compared the ability of the learned embedding to predict the pleomorphism grade and tumor proliferation grade associated to each case of a hold-out test set using multi-class ROC-AUC metrics.

\paragraph{\textbf{Contributions}}
{
\begin{itemize}
	\setlength\itemsep{0pt}
	\item[-] To our knowledge, this is the first time that an auto-encoder is proposed to explicitly disentangle the orientation information in images by learning a 2-part structured representation consisting of rotation-invariant real-valued variables and rotation-equivariant angle variables.
	\item[-] We propose to use \se{2}-structured CNNs to generate latent variables with guaranteed equivariance/invariance properties.
	\item[-] We show that such an unsupervised model can quantify nuclear phenotypical variation in histopathology images and that the learned representation can be used in dowstream analysis to predict slide-level target values.
\end{itemize}
}

\section{Related Work}
\label{relatedWork}

\paragraph{\textbf{Representation Learning for Microscopy Image Analysis}}
Automated quantification of morphological features of single-cell images has been a paradigm for comparing populations of cells in high-throughput studies across microscopy modalities and applications \citep{caicedo2017cellprofiling, carleton2018advances, gupta2019deep}.

In particular, machine learning methods were proposed to learn representation directly from image data.
Transfer learning methods use the internal representation of deep learning models that were trained with an independent dataset and task \citep{cruz2015method, pawlowski2016automating, ando2017improving, bayramoglu2016transfer, hou2016automatic}.
Weakly-supervised and self-supervised models use the representation of deep learning models that are trained with the data at hand but optimized to solve an auxiliary {pretext} task \citep{goldsborough2017cytogan, godinez2018unsupervised, hu2018unsupervised, caicedo2018weakly}.

We argue that methods that rely on training a generative adversarial network and that exploit the inner feature maps of the discriminator network as a representation fall into this category as these feature maps do not necessarily correspond to generative factors.
These methods rely on the hypothesis that deep learning models can learn generic features that will generalize to an external task without further assumptions.
However, such generalization is not guaranteed for a difficult medical-related task whose domain is too different from the task and dataset that were used for the original training of the models.
Also, these methods often rely on subsequent fine-tuning {using} the data at hand, and such a transferred learned representation cannot give any direct visual insight {into} the nature of the individual features.

Generative models and in particular latent variable models, were developed to learn the generative factors of cell images as a representation to be used in downstream tasks. 
These methods are based on (sparse) dictionary learning \citep{das2018sparse}, sparse auto-encoding \citep{xu2014stacked,hou2019sparse}, variational auto-encoding \citep{lafarge2018cellVAE, yang2018autoencoder, schau2019variational}, conditional auto-encoding \citep{johnson2017generative} or other auto-encoding frameworks \citep{murthy2017center, song2017hybrid, feng2018breast, huang2018exclusive}.

\paragraph{\textbf{Other Applications of Representation Learning for Computational Pathology}}
We consider unsupervised representation learning of random patches of digital slides related work: such representations {facilitated the achievement} of downstream tasks \citep{huang2011time, kothari2012biological, arevalo2013hybrid, nayak2013classification, chang2013characterization, cruz2013deep, xu2014deep, zhou2014classification, cruz2015comparative, cruz2015method, chang2015stacked, vu2015histopathological, otalora2015combining, arevalo2015unsupervised, xu2015deep, hou2016patch, kwak2017multiview, sari2018unsupervised, bulten2018unsupervised, wang2018weakly, muhammad2019unsupervised}.
However, when supervision is possible, patch-based representation can be achieved in a end-to-end fashion in a multiple-instance-learning framework for a given task \citep{xu2014deep, liu2016hierarchical, hou2016patch, wang2018weakly, chidester2018discriminative, ilse2018attention, momeni2018deep, combalia2018monte, tomczak2018mil, coudray2018mil, ilse2020MILforHIA}.
In \citep{tellez2019neural}, the authors compared different latent variable models in order to compress WSIs and enable their processing in a single run, and investigated their potential on downstream tasks.
Studies on realistic generation of histopathology images \citep{hou2017unsupervised, hou2019robust, quiros2019pathology, dey2020group} showed that decoder-based generative models can embed the fine-grained morphological structures of tissues in a low-dimensional latent space, which is in line with the motivation of our work.

\paragraph{\textbf{Structured Latent Variable Models}}
Although unsupervised latent variable modeling is appealing, \citet{locatello2018challenging} and \citet{dai2019diagnosing}, supported by the work of \citet{ilse2019diva} showed that learning disentangled generative factors is not possible without constraining the learned representation.
This argument justifies the limited performances of the learned representation of baseline VAE models in downstream tasks.
As a solution, methods were proposed to structure the VAE latent space as a form of inductive prior: hyperspherical latent structure \citep{davidson2018hyperspherical}, supervised nuisance variable separation \citep{louizos2016fair} or domain-wise latent space factorization \citep{ilse2019diva} for example.

In the context of single-cell representation learning, \citet{johnson2017generative} proposed a structured latent space via a conditional VAE model that encourages separation of cell/nuclear shape information from sub-cellular component localization. 

The framework proposed in this paper is in the direction of research of these prior works but specifically address the spatial orientation of the generative factors of cell images.

\paragraph{\textbf{Rotation-Equivariance in Convolutional Networks}}
Deep Learning methods were proposed to learn representations that are equivariant to the orientation of images.
These methods enable learning a representation that changes in a deterministic way when the input image is rotated.
In particular, group convolutional networks \citep{cohen2016group, bekkers2018roto, hoogeboom2018hexaconv, worrall2018cubenet, weiler20183d, cohen2018spherical, cohen2019gauge} {extend} standard CNNs by replacing the convolution operation.
Advantages of using group-structured convolutional networks were shown on computational pathology tasks in a supervised training context \citep{bekkers2018roto, veeling2018rotation, chidester2019nuclear, graham2019rota, lafarge2020roto, graham2020dense}.

Here we leverage the structure of \se{2}-CNNs in an unsupervised context as a new way to structure the latent space of VAE-based models.

\section{Datasets}
\label{datasets}
To train and compare the models investigated in this study, we used {one} dataset for training purposes and to assess slide-level classification performances and {another} patch-based benchmark dataset to assess cell-level classification performances.

\paragraph{\textit{TUPAC-ROI}}
We used a dataset of 148 WSIs of Hematoxylin-Eosin-stained (H{\&}E) tissue slices of breast cancer patients.
These WSIs are part of the training set of the \textit{TUPAC16} challenge \citep{veta2019tupac} and are originally provided by The Cancer Genome Atlas Network \citep{tcga2012}.
Three Regions of Interest (ROIs) were annotated by a pathologist to indicate tumor regions with high cellularity, {that pathologists would typically select for cancer grading}.
Note that we used this subset of WSIs as it was the only subset for which ROIs were provided.

\citet{heng2017molecular} provided several patient-level metrics on these cases, including molecular and genetic information as well as expert-based visual morphological assessment (mitosis grading, tubular formation, pleomorphism grading).
We used the pleomorphism grade and tumor proliferation grade (discrete grades in $\{1, 2, 3\}$) associated to each WSI as a target value to evaluate the quality of the learned representations investigated in this study.
In the case of tumor proliferation grading, we make the assumption that this value can be associated to cell-level patterns and their local distributions.

To reduce the inter-case staining variability, we pre-processed all images by applying the well-established staining normalization method described in \citep{macenko2009method}.
We applied our internal nuclei segmentation deep learning model \citep{lafarge2020roto} within each ROI, and used the center of mass of the segmented instances as an estimate of the nuclei center locations.
Image patches of size $68{\times}68\textrm{px}^{2}$ at a resolution of ${\sim}0.25{\mu}\textrm{m/px}$ centered on these locations were extracted and constitute the {effective} dataset of cell-centered images we used to train and test our models.
We made a training-validation-test split of this dataset (including respectively $104$, $22$ and $22$ cases). {See the supplementary material for details about the class distributions across the splits.}
We will refer to this {refined dataset} as \textit{TUPAC-ROI}.

\paragraph{\textit{CRCHistoPhenotypes}}
In order to assess the single cell-level quality of the trained models and investigate the transfer ability of the learned representation to other tissue types, we used the classification subset of the \textit{CRCHistoPhenotypes - Labeled Cell Nuclei Data} (CRCHP) dataset provided by \citet{sirinukunwattana2016locality}.
This dataset consists of $22,444$ localized nuclei extracted from $100$ ROIs, themselves originating from WSIs of H{\&}E stained histology images of colorectal adenocarcinomas.
Cell-type labels (\textit{epithelial}, \textit{inflammatory}, \textit{fibroblast}, \textit{miscellaneous}) were provided for each nucleus.
We made a training-validation-test split of this dataset free of any ROI-overlap (including respectively $11,090$, $3,133$ and $8,221$ nuclei).
We resampled and cropped image patches centered at the nuclei locations so that the resolution and dimensions matched the ones of the \textit{TUPAC-ROI} dataset.
Finally, we applied the same staining normalization protocol as for the \textit{TUPAC-ROI} dataset.

\section{Methods}
\label{methods}

This section describes the proposed framework, first summarizing the baseline VAE framework, then presenting its expansion using \se{2,N}-group convolutions and finally developing how we enable disentanglement of the orientation information.

We formalize the representation learning problem in this generative modeling setting as the problem of learning the joint distribution $p(\bm{x}, \bm{z})$ of the {observed images} $\bm{x}$ with their latent generative factors $\bm{z}$.
Typically, we want to estimate the distribution that maximizes the marginal likelihood $p(\bm{x})$ of this model for a given dataset.

\subsection{Variational Auto-Encoder}
\label{sec:vae}
In the VAE framework \citep{kingma2013vae}, the likelihood of the observed images given a latent embedding $p_{\psi}(\bm{x} | \bm{z})$ is modeled by a decoder CNN with parameters $\psi$.
It is assumed that the latent $\bm{z}$ are drawn from a given prior distribution $p(\bm{z})$, typically a multivariate normal distribution.
By introducing an approximation of the posterior on the latent $q_{\phi}(\bm{z} | \bm{x})$ modeled by a CNN encoder (parameterized by $\phi$), \citet{kingma2013vae} propose to optimize $\psi$ and $\phi$ by maximizing a tractable lower bound on the marginal log likelihood, as written in Eq.\ref{eqn:vaeObjective}.
\begin{equation}
\label{eqn:vaeObjective}
\resizebox{.9\hsize}{!}{%
$\displaystyle  \mathcal{L}_{\text{\tiny VAE}}(\bm{x}, \bm{z}; \psi, \phi) =
{\mathbb{E}_{q_{\phi}(\bm{z} | \bm{x})}} [ \log p_{\psi}(\bm{x} | \bm{z}) ]
{-} \beta \cdot \textrm{D}_{\scriptscriptstyle \textrm{KL}}\left[q_{\phi}(\bm{z} | \bm{x}) {\ }||{\ } p(\bm{z})) \right]$}
\end{equation}
The Kullback-Leibler divergence term $\textrm{D}_{\scriptscriptstyle \textrm{KL}}$, encourages the distribution of the sampled latents to be close to the prior distribution.
The $\beta$ hyper-pararameter controls the strength of this constraint as introduced by \citet{higgins2017beta}.

\paragraph{Orientation Encoding}
The encoder and decoder CNNs of conventional VAE models are built as a series of alternating trainable 2D convolution operations, non-linearity activation functions and down/up-pooling operations.
For a given image $\bm{x} {\in} \mathbb{L}_{2}[ \mathbb{R}^{2} ]$, the encoder CNN $q_{\phi}$ generates low-dimensional embedding samples $\bm{z} \sim q_{\phi}(\bm{z} | \bm{x})$ (of $M$ elements) with $\bm{z} {=} [ z_{i} ]_{i=1}^{M}$  and $z_{i} \in \mathbb{R}$.

In the context of tissue imaging, we make the hypothesis that every image can be decomposed as a pair $\bm{x} = (\bm{x}_{0}, \theta_{0})$ of independent variables such that $p(\bm{x}) = p(\bm{x}_{0}) {\cdot} p(\theta_{0})$.
With this formulation, we assume the existence of a reference distribution of images $\bm{x}_{0} \sim p(\bm{x}_{0})$ that get rotated by an angle drawn from a uniform distribution $\theta_{0} \sim U(0, 2\pi)$.
This uniformity assumption is logical for histopathology since tissue slices are prepared independently of their orientation and thus tissue images can be acquired in any possible orientation.

We write $\bm{x} = \mathcal{L}_{\theta_{0}}[ \bm{x}_{0} ]$ as the relationship between these variables, with $\mathcal{L}_{\theta_{0}}: \mathbb{L}_{2}[ \mathbb{R}^{2} ] \rightarrow \mathbb{L}_{2}[ \mathbb{R}^{2} ]$ the \textit{left-regular representation} on 2D images of the rotation group \so{2}, parameterized by $\theta_{0}$.
In this notation $\theta_{0}$ indicates the action of a planar rotation $R_{\theta_{0}} {\in} \so{2}$, such that $\mathcal{L}_{\theta_{0}}[ \bm{x}_{0} ](\bm{u}) = \bm{x}_{0}(R^{-1}_{\theta_{0}} {\cdot} \bm{u})$, given a vector location $\bm{u} {\in} \mathbb{R}^{2}$.

From this point of view, $\theta_{0}$ is a generative factor of the observed images, and so it would be expected that the model learns to isolate this factor in the learned latent $\bm{z}$.
Without loss of generality, we can assume that an optimal model would learn to decompose the latent variables such that $\bm{z} = [\bm{z}_{0}, \theta_{0}]$ with $\bm{z}_{0}$ the subset of latent variables that are independent of $\theta_{0}$.

This decomposition would imply that the posterior distribution of $\theta_{0}$ is \textit{equivariant} under the deterministic action of the \textit{rotation group} on the image domain, and via the concatenation of rotations on the domain of orientations (written as the addition of angles modulo $2\pi$ in Eq.~\ref{eqn:vaeEquivariance}).
The desired (conditional) independence relationship between $\bm{z}_{0}$ and $\theta_{0}$ is equivalent to the posterior being \textit{invariant} under the same group actions (see relationship of Eq.~\ref{eqn:vaeInvariance}).

\resizebox{.9\linewidth}{!}{
  \begin{minipage}{\linewidth}

	\begin{subequations}
	\begin{align}
p( \theta_{0} + \theta \spipe \textbf{x} = \mathcal{L}_{\theta}[ \bm{x}_{0} ]) = p( \theta_{0} \spipe \textbf{x} = \bm{x}_{0}) \; {;} \; \forall \theta \in [0, 2\pi) \label{eqn:vaeEquivariance} \\
p( \bm{z}_{0} \spipe \textbf{x} = \mathcal{L}_{\theta}[ \bm{x}_{0} ]) = p( \bm{z}_{0} \spipe \textbf{x} = \bm{x}_{0}) \; {;} \; \forall \theta \in [0, 2\pi) \label{eqn:vaeInvariance}
	\end{align}
\end{subequations}
\end{minipage}}\newline

However empirical experiments have shown that perfect independence of the latent variables is hard to achieve in a generative model and that such disentanglement of the generative factors is typically not obtained without further constraints on the models \citep{higgins2017beta}.
At that, the assumed uniform distribution of $\theta_{0}$ is not encouraged by the Gaussian prior distribution of the formulation of the standard VAE.
Therefore, we propose to consider a network architecture that explicitly encodes the orientation information of the images, guarantees the relationships of Eq.~\ref{eqn:vaeEquivariance}-b and enables the modeling of the Bayesian network illustrated in Figure.~\ref{fig:graphicalModels}-c.

\begin{figure*}[ht!]
\begin{center}
\includegraphics[width=0.9\textwidth, trim=5pt 30pt 60pt 10pt, clip]{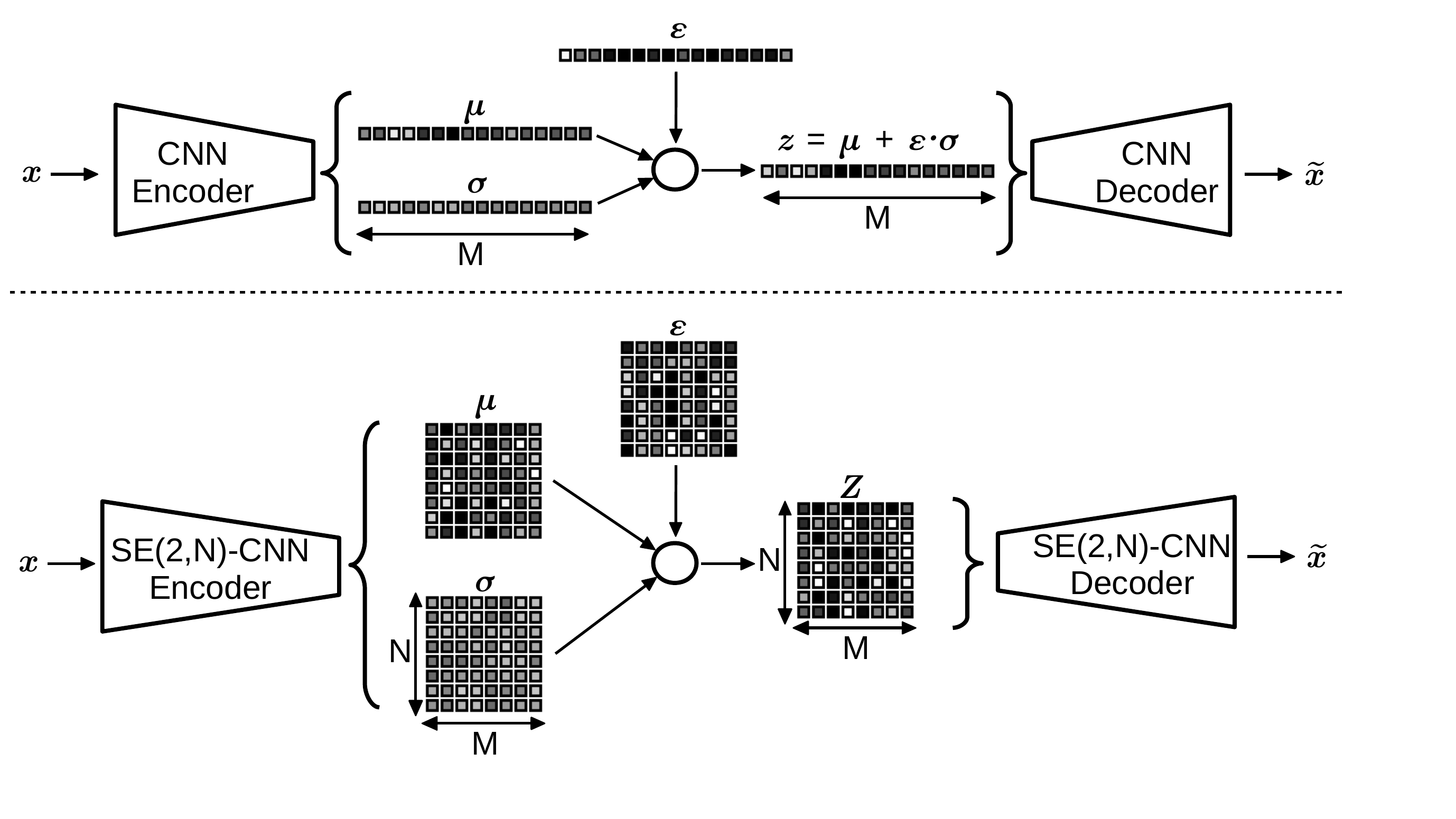}
\caption{\footnotesize
{
Comparison of the auto-encoding pipeline of an image $\bm{x}$ in a conventional VAE (top) and an \se{2,N}-CNN-based VAE (bottom).
Here, $\bm{\mu}$ and $\bm{\sigma}$ parameterize $q_{\phi}(\bm{z} | \bm{x})$ and $q_{\phi}(\bm{Z} | \bm{x})$ such that samples $\bm{z} = [z_{i}]_{i=1 {\ldots} M}$ and $\bm{Z} = [Z_{i,j}]_{i,j=1 {\ldots} M}$ are drawn using the reparametrization trick with $\bm{\epsilon} = [ \epsilon_{i} \sim \mathcal{N}(0,1) ]_{i=1 {\ldots} M}$ or $\bm{\epsilon} = [ \epsilon_{i,j} \sim \mathcal{N}(0,1) ]_{i,j=1 {\ldots} M}$.
}
}
\label{fig:zSamplingVAE}
\end{center}
\end{figure*}

\subsection{\se{2} Variational Auto-Encoder}
\label{sec:se2vae}
Although conventional CNNs are equivariant to translations (since 2D convolutions are equivariant to planar translations), they are not guaranteed to be equivariant with respect to rotations of the input images. 

\paragraph{Group Structured CNNs}
Group convolution operations were proposed to give CNNs the structure of the roto-translation group $\se{2} := \mathbb{R}^{2} {\rtimes} \so{2}$ \citep{cohen2016group}.
The internal feature maps of CNNs with such group structure can be treated as \se{2}-images $F \in \mathbb{L}_{2}[\se{2}]$ and the application of convolutional operations with \se{2}-image kernels are equivariant under the action of the elements of \so{2}.
This architecture provides an end-to-end roto-translation equivariance property to CNNs.

The architecture of a \textit{\se{2}-CNN} can be implemented by discretizing the sub-group $\se{2,N} := \mathbb{R}^{2} {\rtimes} \so{2,N}$ by sampling \so{2} with the elements that correspond to the $N$ rotation angles of $\lbrace {2{\pi}n} / {N} \spipe n{=}{0}{\ldots}{N-1} \rbrace$ \citep{bekkers2018roto, lafarge2020roto}.
This way, the internal feature maps of the network can be implemented as tensors of shape $H {\times} W {\times} N {\times} M$ with $H$ and $W$ the size of the spatial dimensions, $N$ the size of the discretized orientation-axis and $M$ the number of channels in the layer.

\paragraph{Application to the VAE framework}
We propose to replace the 2D-convolution operations of the conventional CNNs in the VAE framework by \se{2}-group convolutions to yield rotation-equivariant mappings from the input images to the sampled latent variables and from the latent variables to the reconstructed images.
This change of architecture also relies on the replacement of the first layer of the encoder by a \textit{lifting layer} to produce \se{2}-image representation maps; and the replacement of the penultimate layer of the decoder by a \textit{projection layer} to output 2D images \citep{bekkers2018roto, lafarge2020roto}.

The bottleneck of a conventional CNN-based encoder corresponds to feature vectors of size $M$.
Likewise, in a \se{2}-CNN-based VAE, the feature vectors at the bottleneck of the encoder can be defined in terms of the rotation elements of the circle group \so{2} solely.
We treat these feature vectors as $\mathcal{U}_{g}[ f ]$ where $\mathcal{U}_{g}$ is the \textit{left-regular group-representation} of \so{2} on functions $f \in \mathbb{L}_{2}[\so{2}]$ with $g = R_{\theta} \in \so{2}$ (we simplify this notation to $\mathcal{U}_{\theta}f$).

In practice, we consider the sub-group \so{2,N}, so that the \so{2,N}-vectors $\mathcal{U}_{\theta}\bm{f}$ can be implemented as tensors of shape $N{\times}M$ with $N$ the size of the discretized orientation-axis and $M$ the number of latent variables.
The difference in embedding structure between conventional VAEs and \se{2}-CNN-based VAEs is illustrated in Figure.~\ref{fig:zSamplingVAE}.

\paragraph{\se{2,N}-Structured Latent Variables}
Instead of considering real-valued latent variables, we propose to model the latent variables as \so{2,N}-vector-valued random variables $\bm{Z}$.
The group structure of the \se{2}-modified encoder enables to model the posterior distribution $q_{\phi}(\bm{Z} \spipe \bm{x})$ with the property of being equivariant under the action of \so{2,N}.
Thus, the modeled distribution verifies the relationship of Eq.~\ref{eqn:se2vaeEquivariance}.
\begin{equation}
\label{eqn:se2vaeEquivariance}
\resizebox{.9\hsize}{!}{%
$q_{\phi}(\mathcal{U}_{\theta}[ \bm{Z} ] \spipe \bm{x} = \mathcal{L}_{\theta}[ \bm{x}_{0} ]) = q_{\phi}(\bm{Z} \spipe \bm{x} = \bm{x}_{0}) {\ } {;} {\ } \forall \theta \in \text{\textit{SO(2,N)}} $}
\end{equation}

The \se{2}-CNN decoder takes the samples $\bm{Z}$ as input and models the likelihood $p_{\psi}(\bm{x} | \bm{Z})$ as a multivariate Gaussian with identity covariance.
Note that this \se{2}-CNN-based VAE can be trained with the same objective as conventional VAEs (see Eq.\ref{eqn:vaeObjective}) after adjusting the prior on the latent to a multivariate normal distribution that matches the dimensions of $\bm{Z}$.

\paragraph{Consequences for Downstream Analysis}
By construction, the equivariance property of the variational posterior $q_{\phi}$ guarantees that rotating the input images by an angle measure $\theta$ will cause a cycle-shift on the posterior distribution as expressed by the relationship of Eq.~\ref{eqn:se2vaeEquivariance}.
Likewise, the rotational equivariance of $p_{\psi}$ implies that cycle-shifting the values of a latent sample will cause a rotation of the reconstructed images.

As a result, the generative process of the images does not depend directly on $\theta_{0}$ anymore: this variable becomes encoded in the \so{2,N} latent variables as a \textit{hidden shift} on the orientation-axis.
Still, the dependence of each variable $Z_{i,j}$ to $\theta_{0}$ makes downstream analysis of these generative factors subject to the variability of this arbitrary orientation within a dataset.

\subsection{Separation of Isotropic and Oriented Latents}
\label{sec:rvae}
In order to achieve disentanglement of the orientation information in the latent variables, we make the hypothesis that the set of generative factors can be split in two sets of independent variables: 
a set of real-valued variables $\bm{z^{\sssize{ISO}}} = [ z^{\sssize{ISO}}_{i} ]_{i=1}^{M}$ that codes for non-oriented/isotropic features in the images, and a set of angle variables $\bm{z^{\sssize{ORI}}} = [ z^{\sssize{ORI}}_{i} ]_{i=1}^{M}$ with values in $\left[0,2\pi\right]$ that code for oriented structures in the images.

To achieve such partitioning of the latent space, we design the \se{2,N}-CNN encoder to approximate two posterior distributions $q_{\phi}(\bm{z}^{\sssize{ISO}} | \bm{x})$ and $q_{\phi}(\bm{z}^{\sssize{ORI}} | \bm{x})$ by producing three output components that parameterize these distributions (as illustrated in Figure.~\ref{fig:zSamplingORIVAE}):
\begin{itemize}
	\setlength\itemsep{0pt}
	\item[-] Two sets of $\so{2,N}$-vectors that are projected via the \textit{mean} operator along the orientation-axis resulting in a \textit{mean vector} $\bm{\mu}^{\sssize{ISO}} \in \mathbb{R}^{M}$ and a \textit{variance vector} $(\bm{\sigma}^{\sssize{ISO}})^{2} \in (\mathbb{R}^{+})^{M}$. 
	\item[-] {One set of \textit{softmax-activated} $\so{2,N}$-vectors $Q^{\sssize{ORI}}_{i}$ that correspond to discretized approximations of $q_{\phi}(z^{\sssize{ORI}}_{i} | \bm{x})$ as defined in Eq.~\ref{eqn:se2oriModeling} with $i=1 {\ldots} M$.
	Here the softmax function is used to ensure that each vector $Q^{\sssize{ORI}}_{i}$ represents {a probability mass function}.}
\end{itemize} 
\begin{equation}
\label{eqn:se2oriModeling}
Q^{\sssize{ORI}}_{i,j} = \int_{\frac{(j-1)2{\pi}}{N}}^{\frac{{j}2{\pi}}{N}} q_{\phi}(z^{\sssize{ORI}}_{i} | \bm{x}) \, \mathrm{d}z^{\sssize{ORI}}_{i} {\ } ; {\ } j=1 {\ldots} N
\end{equation}

Finally, the $z^{\sssize{ISO}}_{i}$ can be drawn from $\mathcal{N}(\mu^{\sssize{ISO}}_{i}, (\sigma^{\sssize{ISO}}_{i})^{2})$ and the $z^{\sssize{ORI}}_{i}$ can be directly drawn from $q_{\phi}(z^{\sssize{ORI}}_{i} | \bm{x})$ using the approximations $Q^{\sssize{ORI}}_{i}$ (the implementation of these sampling procedures are detailed in the next paragraph).

By construction, the $\mu^{\sssize{ISO}}_{i}$ and $\sigma^{\sssize{ISO}}_{i}$ are rotation-invariant, and thus ensure that $\bm{z}^{\sssize{ISO}}$ verifies the \textit{invariance} relationship of Eq.~\ref{eqn:vaeInvariance}.
The modeled posteriors $q_{\phi}(z^{\sssize{ORI}}_{i} | \bm{x})$ follow the equivariance relationship of Eq.~\ref{eqn:vaeEquivariance}, as $\theta_{0}$ becomes encoded as a shared hidden shift (modulo $2\pi$) across the variables $z^{\sssize{ORI}}_{i}$.

\begin{figure*}[ht!]
\begin{center}
\includegraphics[width=0.9\textwidth, trim=5pt 5pt 125pt 5pt, clip]{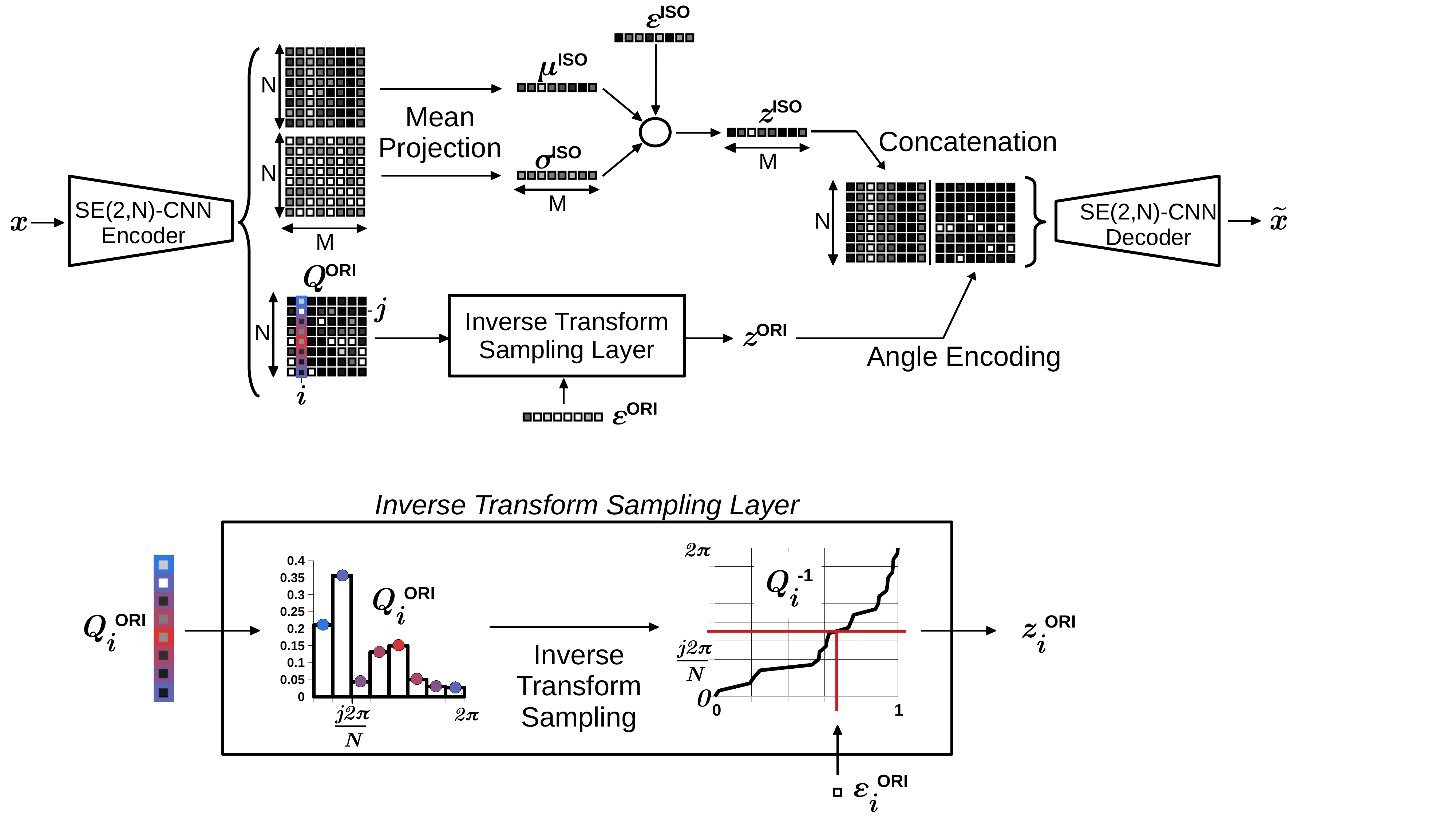}
\caption{\footnotesize
{
Illustration of the auto-encoding pipeline of an image $\bm{x}$ in the proposed VAE.
Here, the embedding consists of two components $\bm{z}^{\sssize{ISO}}$ and $\bm{z}^{\sssize{ORI}}$.
$\bm{z}^{\sssize{ISO}}$ is drawn from $q_{\phi}(\bm{z}^{\sssize{ISO}} | \bm{x})$ parameterized by $\bm{\mu}^{\sssize{ISO}}$ and $\bm{\sigma}^{\sssize{ISO}}$ such that samples $\bm{z}^{\sssize{ISO}} = [z^{\sssize{ISO}}_{i}]_{i=1 {\ldots} M}$ are drawn using the reparametrization trick with $\bm{\epsilon}^{\sssize{ISO}} = [ \epsilon^{\sssize{ISO}}_{i} \sim \mathcal{N}(0,1) ]_{i=1 {\ldots} M}$.
$\bm{z}^{\sssize{ORI}}$ is drawn from $q_{\phi}(\bm{z}^{\sssize{ORI}} | \bm{x})$ via Inverse Transform Sampling applied on the cumulative density functions approximated {by} $\bm{Q}^{\sssize{ORI}}$ using a noise variable $\bm{\epsilon}^{\sssize{ISO}} = [ \epsilon^{\sssize{ORI}}_{i} \sim \mathcal{U}(0,1) ]_{i=1 {\ldots} M'}$ (an example of this sampling process is shown for a variable $\bm{z}_{i}^{\sssize{ORI}}$).
}
}
\label{fig:zSamplingORIVAE}
\end{center}
\end{figure*}

\paragraph{Implementation Details on Latent Sampling}
During training, the stochastic sampling process of the $\bm{z}^{\sssize{ISO}}$ and $\bm{z}^{\sssize{ORI}}$ variables requires implementation via differentiable operations in the computational graph, to enable gradient back-propagation through the encoder.

As proposed by \citet{kingma2013vae}, we implemented the sampling process of $\bm{z}^{\sssize{ISO}}$ via the reparameterization trick to obtain sampled latents $\bm{z}^{\sssize{ISO}} = \mu^{\sssize{ISO}} + \epsilon^{\sssize{ISO}} {\cdot} \sigma^{\sssize{ISO}}$ (as illustrated in Figure.~\ref{fig:zSamplingVAE}).
Here, $\epsilon^{\sssize{ISO}} \sim \mathcal{N}(0,1)$ is an auxiliary noise variable to simulate sampling from a Gaussian distribution.

To approximate the sampling process of $z^{\sssize{ORI}}$ we implemented an \textit{Inverse Transform Sampling} layer that returns a set of angle measures drawn from the distributions coded by $Q^{\sssize{ORI}}$.
This layer computes a \textit{continuous inverse cumulative distribution} $Q^{\sssize -1}_{i}$ for each vector $Q^{\sssize{ORI}}_{i}$ and calculates each angle measure as $\bm{z}^{\sssize{ORI}} = Q^{\sssize -1}_{i}(\epsilon^{\sssize{ORI}})$ with $\epsilon^{\sssize{ORI}} \sim \mathcal{U}(0, 1)$ a uniformly distributed auxiliary noise variable.
An example of this two-step procedure is shown in Figure.~\ref{fig:zSamplingORIVAE}.

So as to conserve the end-to-end equivariance property of the framework, the latent variables are reshaped as \se{2,N}-vectors, such that the decoding procedure can occur in the same conditions as described in Section.~\ref{sec:se2vae}.
This is done by expanding and repeating the values of $\bm{z}^{\sssize{ISO}}$ along the orientation-axis,
and via \textit{label smoothing} to encode the sampled angles $\bm{z}^{\sssize{ORI}}_{i}$ into \se{2,N}-vectors.

\paragraph{Extended Objective}
Since the images are generated from two independent sources of generative factors we simply further developed the lower bound in the VAE formulation such that the proposed model can be trained by maximizing the loss written in Eq.~\ref{eqn:vaeIsoOriObjective}.
The constraints are computed by fixing the prior $p(\bm{z}^{\sssize{ISO}})$ as a multivariate normal distribution and the prior $p(\bm{z}^{\sssize{ORI}})$ as a multivariate uniform distribution on $\left[0, 2\pi \right]$.

\resizebox{.9\linewidth}{!}{
  \begin{minipage}{\linewidth}
	\begin{align}
\label{eqn:vaeIsoOriObjective}
\mathcal{L}_{\theta \textrm{-VAE}}(\bm{x}, \bm{z}^{\sssize{ISO}}, \bm{z}^{\sssize{ORI}}; \psi, \phi) {\ }{=}{\ } &
{ \mathbb{E}_{q_{\phi}(\bm{z}^{\sssize{ISO}}, \bm{z}^{\sssize{ORI}} | \bm{x})} } \left[ \log p_{\psi}(\bm{x} \spipe \bm{z}^{\sssize{ISO}}, \bm{z}^{\sssize{ORI}}) \right] \notag \\
 & {-} \beta^{\sssize{ISO}} \cdot \textrm{D}_{\scriptscriptstyle \textrm{KL}}\left[q_{\phi}(\bm{z}^{\sssize{ISO}} | \bm{x}) {\ }||{\ } p(\bm{z}^{\sssize{ISO}}) \right] \notag \\
  & {-} \beta^{\sssize{ORI}} \cdot \textrm{D}_{\scriptscriptstyle \textrm{KL}}\left[q_{\phi}(\bm{z}^{\sssize{ORI}} | \bm{x}) {\ }||{\ } p(\bm{z}^{\sssize{ORI}}) \right]
	\end{align}
   \end{minipage}}

\paragraph{Consequences for Downstream Analysis}
The isotropic generative factors $\bm{z^{\sssize{ISO}}}$ are guaranteed to be independent to $\theta_{0}$ by construction, so they can be compared and aggregated within/across populations of tissue image patches independently of their spatial orientation.
Likewise, the distribution of angles $z^{\sssize{ORI}}$ in a given population characterizes the variability of oriented features independently of isotropic factors.

\subsubsection{Complementary Reconstruction Loss}
Conventional VAE models are known for generating/reconstructing \textit{blurry} images of a lower quality than original images.
Poor reconstructions might imply that the high-frequency details in the images do not get encoded in the latent representation and might entail poor performances in downstream tasks.
This limited quality of generated images is often associated to the pixel-wise reconstruction term of the VAE objective (see Eq.~\ref{eqn:vaeObjective}).

To ensure the reconstruction term {enables} generation of realistic images, we used the extension of the VAE objective that was described in \citep{lafarge2018cellVAE}.
This method is {inspired} by the initial work of \citet{larsen2016vaegan} {that introduces} a discriminator CNN $D$ with parameters $\chi$ in the VAE framework.

This discriminator is trained to classify batches balanced between real images $\bm{x} {\sim} p(\bm{x})$ and reconstructed images $\bm{\widetilde{x}} {\sim} p_{\psi}(\bm{x} | \bm{z})$.
The internal feature maps of $D$ for a given input image $\bm{x}$ are defined as $D_{i}(\bm{x})$ with $i$ a layer index.
The $D_{i}(\bm{\widetilde{x}})$ can be considered as additional Gaussian observations with identity covariance drawn from $p_{\chi}(D_{i}(\bm{x}) | \bm{z})$.
We thus define extra reconstruction losses $\mathcal{L}^{D}_{i}$ that we use to complement the model objective as defined in Eq.~\ref{eqn:vaePlusObjective} where $\gamma$ is a weighting hyper-parameter.

\resizebox{.9\linewidth}{!}{
  \begin{minipage}{\linewidth}
	\begin{align}
\label{eqn:vaePlusObjective}
& \mathcal{L}_{\text{\tiny VAE}+}(\bm{x}, \bm{z}; \psi, \phi, \chi) {\ }{=}{\ } 
\mathcal{L}_{\text{\tiny VAE}}(\bm{x}, \bm{z}; \psi, \phi) + \gamma \sum_{i} \mathcal{L}^{D}_{i}(\bm{x}, \bm{z}; \psi, \phi, \chi) \notag \\
& \mathcal{L}^{D}_{i}(\bm{x}, \bm{z}; \psi, \phi, \chi) = \mathbb{E}_{q_{\phi}(\bm{z}|\bm{x})}[ \log p_{\chi}( D_{i}(\bm{x}) \spipe \bm{z} ) ]
	\end{align}
   \end{minipage}}\newline

The training procedure of the full model pipeline {consists of} alternating training {steps} between updating $\psi$ and $\phi$ by maximizing $\mathcal{L}_{\text{\tiny VAE}+}$ and updating $\chi$ by minimizing the cross-entropy loss on the predictions of $D$.

This extension of the VAE framework has the benefit to keep the rest of the model formulation intact and was shown to be beneficial in autoencoder-based applications involving {cell images} \citep{johnson2017generative, lafarge2018cellVAE}.
The discriminator is used at training time only and is discarded at inference time, which thus does not cause any computational overhead for downstream analysis.
All the models investigated in this paper were extended by this method.

\section{Experiments}
\label{experiments}
In this section we detail the network architectures we designed to implement the proposed orientation-disentangled VAE (Section.~
\ref{sec:rvae}) and a comparable baseline VAE model (Section.~\ref{sec:vae}).
We also describe their training procedures and the evaluation protocols we applied to compare the quality of the learned representation and gain insights {into} the effect of the disentanglement and newly introduced hyper-parameters.

\subsection{Model Architectures}
We designed the conventional CNN encoders and SE(2,N)-CNNs encoders of the VAEs as {straight-forward sequences of four blocks that each consists of a convolutional layer, a batch normalization layer (BN), a leaky reLU non-linearity and a max-pooling layer}.

In the case of the \se{2,N}-CNNs, conventional $\mathbb{R}^{2}$-convolutional layers were replaced by \se{2,N} convolutional layers and BN layers were adjusted to include the orientation-axis in the computation of the batch statistics.
{We also introduced intermediate \textit{projection layers} similar to the locally rotation-invariant CNNs proposed by \citet{andrearczyk2019pulmonary} within all the hidden layers in order to reduce the computational cost of the network.}

In order to generate embedding samples at the bottleneck of the models during training, we implemented the computational sampling procedures presented in Section.~\ref{sec:vae} for the conventional VAEs and in Section.\ref{sec:rvae} for the orientation-disentangled VAEs.

The decoder networks correspond to mirrored versions of their encoder counterparts via the use of up-sampling layers and transposed convolutions.
We included a mean-projection layer at the end of the \se{2,N}-CNN decoder to project the features maps on $\mathbb{R}^{2}$.
Finally we added an extra $1{\times}1$-convolutional layer to the decoders to output images whose dimensions match the input dimensions.

For all the networks, we used kernels with spatial size $5{\times}5$ so as to enable proper rotation of the \se{2,N}-kernels as described in \citep{bekkers2018roto, lafarge2020roto}.
We fixed the angular resolution of the \se{2,N} layers to $N{=}8$ as it {was previously shown} to give optimal performances \citep{lafarge2020roto}.
To ensure fair comparison of the models, we balanced the number of channels in each layer such that the total number of weights between the two types of VAEs is approximately the same.
{In order to have a fair comparison,}, we also fixed the total number of variables in the bottleneck of all the models (size of $\bm{z}$ is $64$ in the conventional VAEs and $\bm{z}^{\sssize{ISO}}$ and $\bm{z}^{\sssize{ORI}}$ are both of size $32$ in the orientation-disentangled VAEs).

More details about the model architectures can be found in the Appendix.

\subsection{Training Procedures}
All the models were trained using the training set of the \textit{TUPAC-ROI dataset} described in Section.~\ref{datasets}.
We used mini-batches that consist of $35$ image patches, such that the distribution of the WSIs of origin within each batch was approximately uniform.
We used the \textit{Adam} optimizer to update the weights of the encoders and decoders (learning rate $0.001$, $\beta_{1}{=}0.9$, $\beta_{2}=0.999$), and \textit{Stochastic Gradient Descent} with momentum to update the weights of the discriminator (learning rate $0.001$, momentum $0.9$).
All convolutional kernels were regularized via decoupled weight decay with coefficient $1{\times}10^{-4}$.
We stopped the training process after convergence of the loss on the validation set (approximately $20{\times}10^{3}$ iterations).
The weighting coefficient of the disciminator-based reconstruction loss was fixed to $\gamma{=}0.01$ across all experiments.

In order to assess the effect of the weighting coefficient of the prior constraint as it was evidenced by \citet{higgins2017beta} we trained the models with varying values of $\beta$, $\beta^{\sssize{ISO}}$ and $\beta^{\sssize{ORI}}$ in $\lbrace 0.1, 0.5, 1.0, 2.0, 4.0 \rbrace$.

\subsection{Downstream Analysis}
In order to compare the quality of the learned representation at the bottleneck of the different trained {VAE models}, we trained subsequent logistic regression {models} that take the learned representation as input to solve downstream tasks.

We investigated two types of downstream tasks to evaluate the value of the learned representation in different contexts: patient-level classification tasks (predicting the pleomorphism grade and tumor proliferation grade of a given WSI) using the \textit{TUPAC-ROI} dataset and single-cell-level classification tasks (predicting the cell-type of a given nucleus) using the \textit{CRCHP} dataset.

Each downstream task was investigated with different types of representation of the {data points} based on the different latent variables estimated in the VAEs (see $\bm{z}$ in Section.~\ref{sec:vae}, $\bm{z}^{\sssize{ISO}}$ and $\bm{z}^{\sssize{ORI}}$ in Section.~\ref{sec:rvae}.
{For comparison,} we also considered well-established nuclear morphometric measurements for comparison (mean nuclear area and standard deviation of the nuclear area) as well as combinations of representations.

\paragraph{Patient-level classification tasks:} we first aggregated the representation of all nucleus-centered image patches within every ROI of the dataset (see Section.~\ref{datasets}).
For the representations corresponding to $\bm{z}$ and $\bm{z}^{\sssize{ISO}}$, we considered the mean $\bm{\mu}$ and $\bm{\mu}^{\sssize{ISO}}$ of the estimated posterior distributions as cell-level embeddings, and we further computed ROI-level embeddings by computing the mean of all the embeddings obtained within a ROI.
We then trained a single-layered multi-class logistic regression model to minimize the cross-entropy loss given the WSI-level ground-truth labels.

For the representation corresponding to $\bm{z}^{\sssize{ORI}}$, we used their discrete distribution at the ROI-level as an aggregated representation (see $Q^{\sssize{ORI}}$ in Section.~\ref{sec:rvae}).
Likewise, we trained a two-layer multi-class logistic regression model with an intermediate maximum projection layer to ensure rotational-invariance of the predictions.

\paragraph{Single-cell-level classification task:} we directly used {the mean $\bm{\mu}$ and $\bm{\mu}^{\sssize{ISO}}$ of the posterior distributions} on $\bm{z}$ and $\bm{z}^{\sssize{ISO}}$ as representation of nucleus-centered image patches and trained a single-layered multi-class logistic regression model to minimize the cross-entropy loss given cell-type ground-truth labels.

All logistic regression models were regularized with L\textsubscript{2}-weight decay and the associated coefficient was fine-tuned on the validation set.

\begin{figure*}
\begin{center}
\includegraphics[width=1\textwidth, trim=5pt 200pt 25pt 50pt, clip]{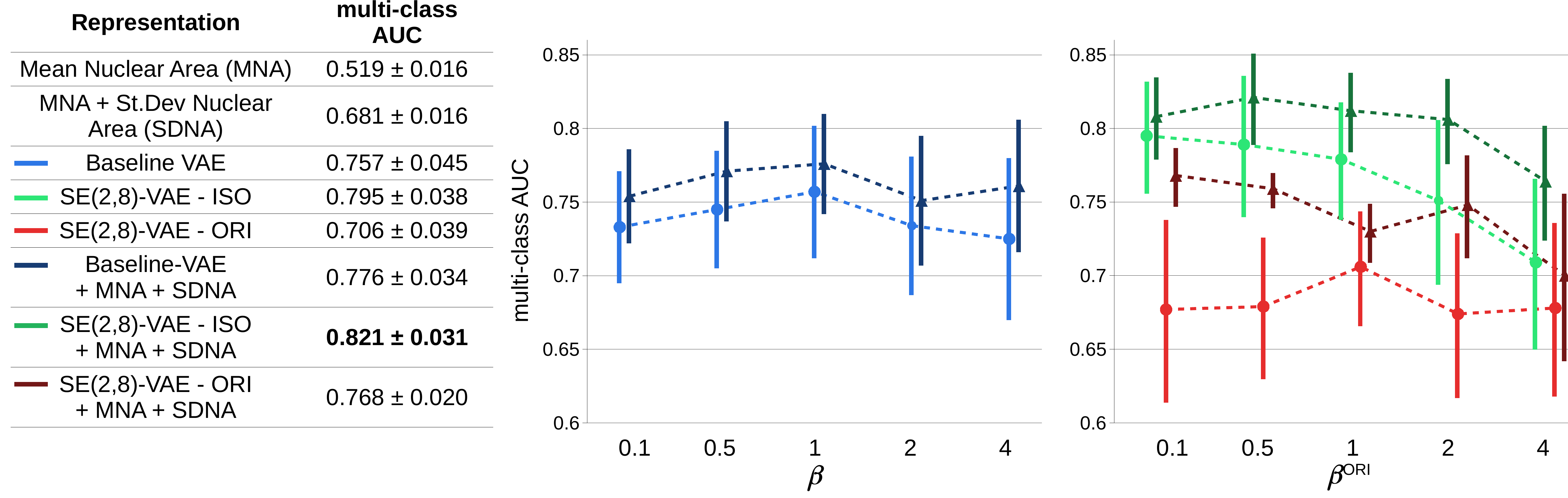}
\caption{\footnotesize{
Performances in downstream analysis for pleomorphism grading.
The table shows best obtained scores for each type of investigated representation.
The plots shows the effect of different hyper-parameters: $\beta$ for the baseline VAE, and $\beta^{\sssize{ORI}}$ with fixed $\beta^{\sssize{ISO}}=1$ for the proposed orientation-disentangled VAE.
Mean $\pm$ standard deviation of the multi-class AUC are indicated in the table and shown with a bar in the plots.
}}
\label{fig:results_pleomorphism}
\end{center}
\end{figure*}

\begin{figure*}
\begin{center}
\includegraphics[width=1\textwidth, trim=5pt 200pt 25pt 50pt, clip]{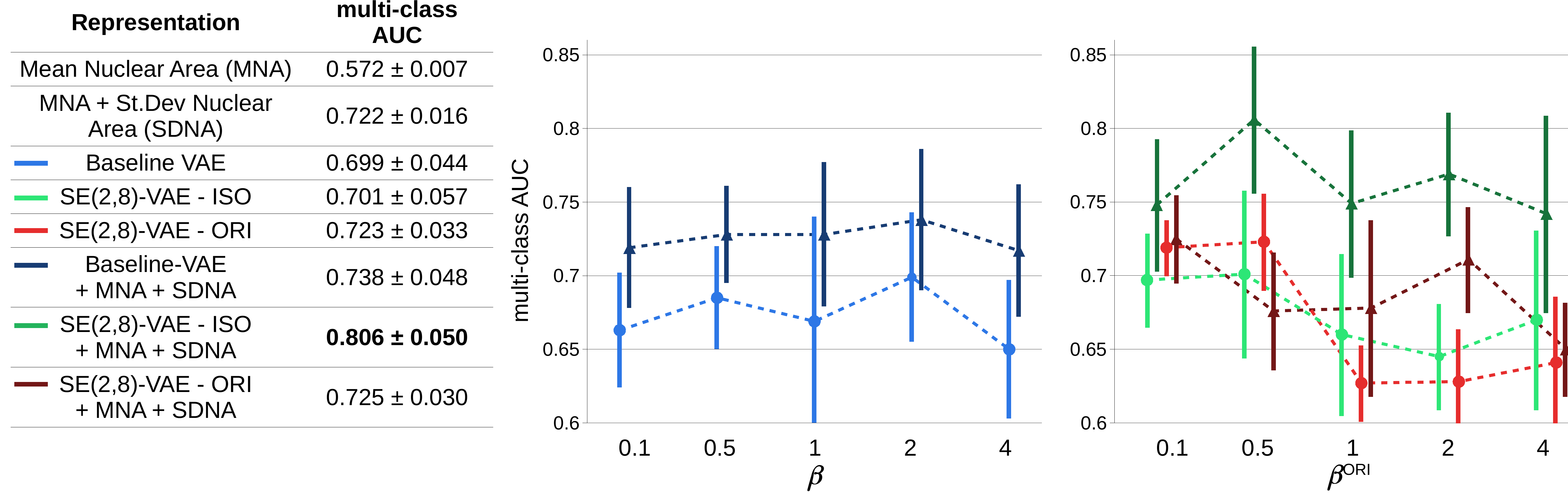}
\caption{\footnotesize{
Performances in downstream analysis for tumor proliferation grade prediction.
The table shows best obtained scores for each type of investigated representation.
Curves shows the effect of different hyper-parameters: $\beta$ for the baseline VAE, and $\beta^{\sssize{ORI}}$ with fixed $\beta^{\sssize{ISO}}=1$ for the proposed orientation-disentangled VAE.
Mean $\pm$ standard deviation of the multi-class AUC are indicated in the table and shown with a bar in the plots.
}}
\label{fig:results_mitosis}
\end{center}
\end{figure*}

\begin{figure*}
\begin{center}
\includegraphics[width=1\textwidth, trim=5pt 330pt 25pt 50pt, clip]{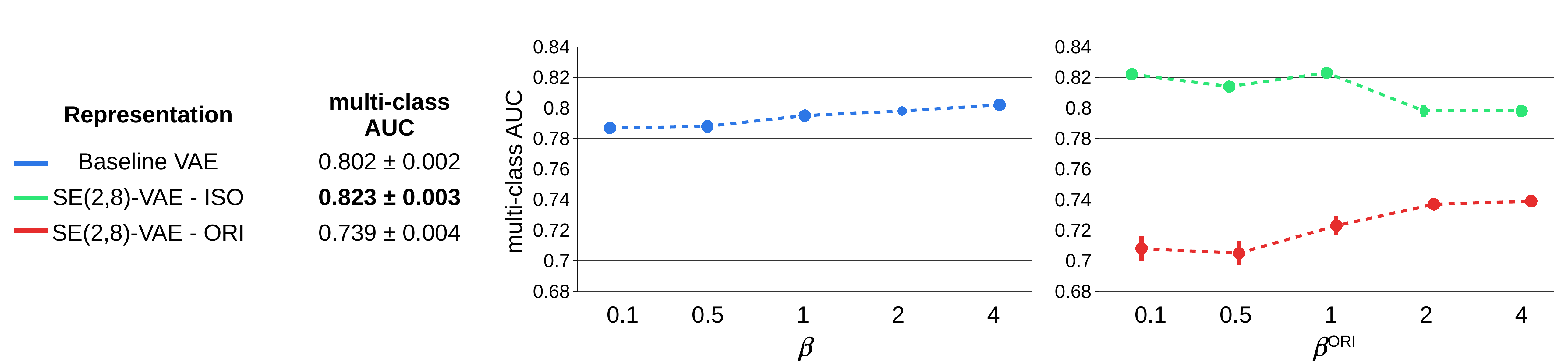}
\caption{\footnotesize{
Performances in downstream analysis for single-cell type classification.
The table shows best obtained scores for each type of investigated representation.
Curves shows the effect of different hyper-parameters: $\beta$ for the baseline VAE, and $\beta^{\sssize{ORI}}$ with fixed $\beta^{\sssize{ISO}}=1$ for the proposed orientation-disentangled VAE.
Mean $\pm$ standard deviation of the multi-class AUC are indicated in the table and shown with a bar in the plots.
}}
\label{fig:results_cellTypes}
\end{center}
\end{figure*}

\section{Results}
\label{results}
This section details the evaluation protocols and metrics we used {to assess the quality of the learned representation we obtained} with the methods presented in Section.~\ref{methods}.
We present the downstream performances for each type of representation on the tasks described in Section.~\ref{experiments}.

\subsection{Slide-Level Downstream Performances}
We evaluated the trained multi-class logistic regression models on the hold-out test set described in Section.~\ref{datasets}.
For each model, we considered a set of binary classifiers that correspond to pairwise comparisons of each class against the other classes.
For each set of such binary classifiers, we computed a set of \textit{Receiving Operating Curves} (ROCs) based on the model predictions.
Within each set of ROCs, we computed the corresponding \textit{Areas Under the Curve} (AUC) and the \textit{mean of AUCs} (mAUC) across that set to summarize the AUC metric given the multi-class setting at hand.

To assess the robustness of the learned representations given perturbations of the training data, {we resampled the training set ten times}, re-trained the models and reported the mean and standard deviation of the mAUCs across these repeats.

\paragraph{Pleomorphism Grading}
The results for the pleomorphism grade prediction task are summarized in Figure.~\ref{fig:results_pleomorphism}.
We report an improvement of the mAUC of $0.038$ from the isotropic learned representation in comparison to the learned representation of the baseline VAE.
The isotropic learned representation {performed better than the oriented learned representation for all the tested values of $\beta^{\sssize{ISO}}$.}
We obtained a consistent additional improvement of performance when combining the isotropic learned representation with segmentation-based features.

\paragraph{Tumor Proliferation Grade Prediction}
The results for the tumor proliferation grade prediction task are summarized in Figure.~\ref{fig:results_mitosis}.
We do not report any significant improvement of the mAUC from using isotropic or oriented learned representation in comparison to the learned representation of the baseline VAE and to segmentation-based features.
{We report} a consistent additional improvement of performance when combining the isotropic learned representation with segmentation-based features.

\subsection{Cell-Level Downstream Performances}
We used the same protocol to evaluate the cell-type classification models.
The results for the cell-type classification task are summarized in Figure.~\ref{fig:results_cellTypes}.
We report a consistent improvement of the mAUC from using the isotropic representation in comparison to the learned representation of the baseline VAE.
The oriented learned representation {performed worse than the representation of the baseline VAE or {the isotropic learned representation} for all the tested values of $\beta$ and $\beta^{\sssize{ISO}}$.}

\section{Discussion and Conclusions}
\label{conclusions}
In this study, we proposed a novel rotation-equivariant variational auto-encoder framework that learns two types of generative factors: isotropic real-valued components and oriented angular components.
We showed that this two-fold low-dimensional structure can efficiently represent histopathology images.
We investigated in a controlled experimental setup, the predictive power of the learned representation using the proposed frameworks and show its advantage in comparison to unsupervised baseline counterparts.
The difference of generative action of each type of variable was qualitatively demonstrated via smooth transitions of generated examples given interpolated embeddings (see Figures.~\ref{fig:generativeProcess}-\ref{fig:generativeInterpolation}).

Qualitatively, we observed that the isotropic learned representation captures morphological factors such as stain ratios, nuclear sparsity, thickness of the nuclear boundary (see Figure.~\ref{fig:generativeProcess}(b)) whereas the oriented learned representation codes for the radial location of the surrounding objects (non-centered neighboring nuclei) and asymmetric structures (see Figure.~\ref{fig:generativeProcess}(a)).

Quantitatively, using isotropic representation was always better or as good as the representation learned by conventional VAE or segmentation-based nuclear area features (see Section.~\ref{results}).
This is in agreement with our hypothesis that the orientation information that is entangled in the representation learned by the baseline VAE affects the quality of the aggregated representation, and subsequent downstream performances.
This was both observed for patient-level grading tasks based on aggregated representation and for cell-level classification tasks.
This is also in line with previous results reported in studies on supervised rotation-invariant CNN models trained to solve computational pathology tasks \citep{bekkers2018roto, veeling2018rotation, chidester2019nuclear, graham2019rota, lafarge2020roto, graham2020dense}.

Although our results suggest that unsupervised learned representation is an efficient alternative to hand-crafted feature-based representations, the fact that the combination of hand-crafted nuclear feature representation with unsupervised learned representation gave a consistent improvement of performances {also reveals the limitations} of the proposed framework.
Indeed, complex knowledge-based quantities such as the \textit{mean nuclear area} that were relevant for the task at hand could not be extracted from the learned representation as they do not necessarily correspond to independent generative factors that the models learned.
However we conjecture that this limitation might be {due to the restricted architectural design of the model} that we chose {for} the comparative analysis.
We thus believe that this limitation can be potentially overcome by using more complex and sophisticated architectures for the latent variable models and downstream classification models.

The performances achieved on the pleomorphism grading task using the isotropic learned representation indicate that these features were more predictive than {the oriented features} to solve this specific task.
This is expected as by definition the pleomorphism grade was labeled based on rotation-invariant nuclear morphological factors.
However on the tumor proliferation grading task, oriented features were slightly more predictive than isotropic features.
This suggests that these two types of features were equally informative of {this specific} patient-level grade.

Besides these applications, the proposed framework can be used as a generic tool to quickly gain insights, and with minimal training effort, into the slide-level predictive value of fixed-scale image patches.
Indeed, we showed that logistic regression models could be trained using aggregated representation of cell populations to predict patient-level target values such as the pleomorphism grade and tumor proliferation grade.
But the same approach could potentially be applied to estimate any other slide-level value.
Also, extensions to other convolutional latent variable models are possible including more complex architectures (such as flow-based generative models), other families of variational distributions (such as structured posterior distributions of the oriented variables) and other training paradigms (such as in a semi-supervised framework).
For computational convenience, we proposed implementing the sampling of the angle variables by means of a straight-forward \textit{inverse sampling layer}, however, other end-to-end sampling strategies could be investigated for further improvement.
This framework {is also transferable} to other problems in which one wants to model a posterior distribution on the rotation group.
Other interesting applications for future work include pre-training for patch-based classification tasks, and compression of WSIs.

\section*{References}
\vspace*{20pt}
\bibliographystyle{unsrtnat}
\renewcommand\refname{}
\vspace*{-40pt}
\bibliography{main}
\balance

\newpage
\includepdf[pages=1-4]{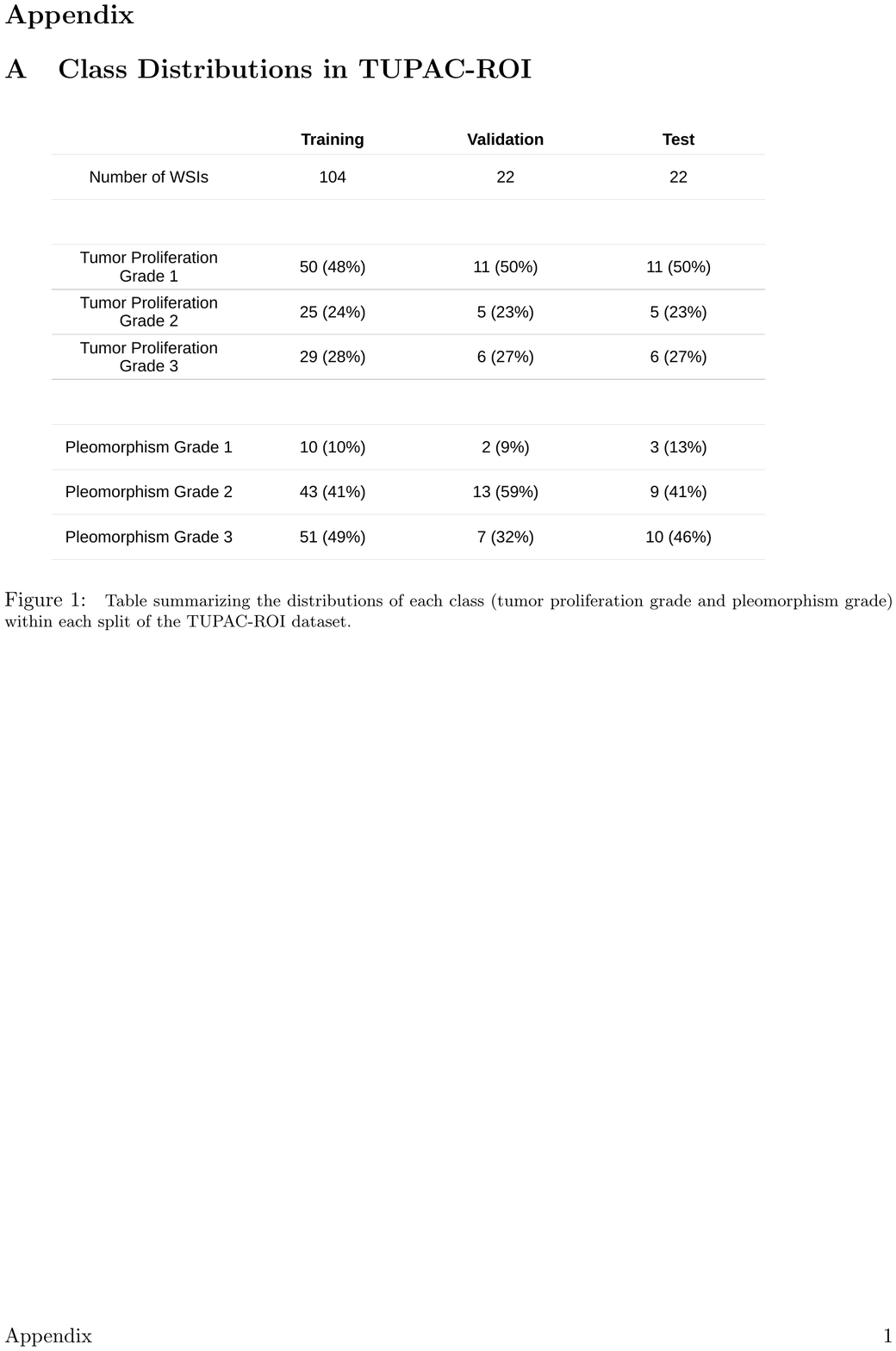}

\end{document}